\documentclass[english,aps,superscriptaddress,preprintnumbers]{revtex4}
\usepackage[T1]{fontenc}
\usepackage[latin9]{inputenc}
\usepackage{textcomp}
\usepackage{mathrsfs}
\usepackage{amsmath}
\usepackage{amssymb}
\usepackage{graphicx}
\usepackage{esint}

\makeatletter
\@ifundefined{textcolor}{}
{%
 \definecolor{BLACK}{gray}{0}
 \definecolor{WHITE}{gray}{1}
 \definecolor{RED}{rgb}{1,0,0}
 \definecolor{GREEN}{rgb}{0,1,0}
 \definecolor{BLUE}{rgb}{0,0,1}
 \definecolor{CYAN}{cmyk}{1,0,0,0}
 \definecolor{MAGENTA}{cmyk}{0,1,0,0}
 \definecolor{YELLOW}{cmyk}{0,0,1,0}
}

\usepackage{mathrsfs}

\@ifundefined{textcolor}{}{%
 \definecolor{BLACK}{gray}{0}
 \definecolor{WHITE}{gray}{1}
 \definecolor{RED}{rgb}{1,0,0}
 \definecolor{GREEN}{rgb}{0,1,0}
 \definecolor{BLUE}{rgb}{0,0,1}
 \definecolor{CYAN}{cmyk}{1,0,0,0}
 \definecolor{MAGENTA}{cmyk}{0,1,0,0}
 \definecolor{YELLOW}{cmyk}{0,0,1,0}
}

\usepackage{babel}

\makeatother

\usepackage{babel}
\begin{document}

\title{Pseudoscalar condensation induced by chiral anomaly and vorticity
for massive fermions}

\author{Ren-hong Fang}

\affiliation{Interdisciplinary Center for Theoretical Study and Department of
Modern Physics, University of Science and Technology of China, Hefei,
Anhui 230026, China}

\author{Jin-yi Pang}

\affiliation{Helmholtz-Institut fuer Strahlen- und Kernphysik (Theorie) and Bethe
Center for Theoretical Physics, Universitaet Bonn, D-53115 Bonn, Germany}

\author{Qun Wang}

\affiliation{Interdisciplinary Center for Theoretical Study and Department of
Modern Physics, University of Science and Technology of China, Hefei,
Anhui 230026, China}

\author{Xin-nian Wang}

\affiliation{Key Laboratory of Quark and Lepton Physics (MOE) and Institute of
Particle Physics, Central China Normal University, Wuhan, 430079,
China}

\affiliation{Nuclear Science Division, MS 70R0319, Lawrence Berkeley National
Laboratory, Berkeley, California 94720}
\begin{abstract}
We derive the pseudoscalar condensate induced by anomaly and vorticity
from the Wigner function for massive fermions in homogeneous electromagnetic
fields. It has an anomaly term and a force-vorticity coupling term.
As a mass effect, the pseudoscalar condensate is linearly proportional
to the fermion mass in small mass expansion. By a generalization to
two-flavor and three-flavor cases, the neutral pion and eta meson
condensates are calculated from the Wigner function and have anomaly
parts as well as force-vorticity parts, in which the anomaly part
of the neutral pion condensate is consistent to the previous result.
We also discuss about possible observables of the condensates in heavy
ion collisions such as collective flows of neutral pions and eta mesons
which may be influenced by the electromagnetic field and vorticity
profiles. 
\end{abstract}
\maketitle
\preprint{\hfill {\small {ICTS-USTC-16-19}}}

\section{Introduction}

The chiral or axial anomaly is the anomalous nonconservation of a
chiral or axial current of fermions arising from the quantum tunneling
effect between vacuums of gauge fields with different winding numbers.
Chiral anomaly is also called Adler-Bell-Jackiw (ABJ) anomaly after
the names of three founders \cite{Adler:1969gk,Bell:1969ts}. In quantum
electrodynamics the anomalous nonconservation of the chiral or axial
current can be written as 
\begin{equation}
\partial_{\mu}j_{5}^{\mu}=-2mP-\frac{Q^{2}}{8\pi^{2}}F_{\mu\nu}\tilde{F}^{\mu\nu},\label{eq:anomaly-2}
\end{equation}
where $m$ and $Q$ are the fermion mass and charge respectively,
$F_{\mu\nu}$ is the strength tensor of electromagnetic field with
$\tilde{F}^{\rho\lambda}=\frac{1}{2}\epsilon^{\rho\lambda\mu\nu}F_{\mu\nu}$
being its dual, chiral or axial current is defined by $j_{5}^{\mu}=\bar{\psi}\gamma^{\mu}\gamma_{5}\psi$,
the pseudoscalar is defined by $P=-i\bar{\psi}\gamma_{5}\psi$, where
$\psi$ and $\bar{\psi}$ are fermionic fields, $\gamma^{\mu}$ $(\mu=0,1,2,3)$
are Dirac matrices and $\gamma^{5}=i\gamma^{0}\gamma^{1}\gamma^{2}\gamma^{3}$
is the chiral matrix. The most successful test of chiral anomaly is
in the decay of a neutral pion into two photons, which had been a
puzzle for some time in 1960s whose solution led to the discovery
of the ABJ anomaly. For neutral pions, one can define the chiral current
as $j_{5,\pi}^{\mu}=\bar{\psi}\gamma^{\mu}\gamma_{5}(\sigma_{3}/2)\psi$
and pseudoscalar as $P_{\pi}=-i\bar{\psi}\gamma_{5}(\sigma_{3}/2)\psi$,
where $\psi=(u,d)^{T}$ and $\bar{\psi}=(\bar{u},\bar{d})$ are quark
fields of two flavors and $\sigma_{3}=\mathrm{diag}(1,-1)$ is the
third Pauli matrix. The anomaly equation (\ref{eq:anomaly-2}) now
becomes 
\begin{equation}
\partial_{\mu}j_{5,\pi}^{\mu}=f_{\pi}m_{\pi}^{2}\phi_{\pi}-\frac{Q_{e}^{2}}{16\pi^{2}}F_{\mu\nu}\tilde{F}^{\mu\nu},\label{eq:pcac}
\end{equation}
where $Q_{e}$ is the absolute value of the electron's charge. Here
we have used the PCAC (partially conserved axial current) hypothesis
\cite{Adler:1964um} to relate the pseudoscalar $P_{\pi}$ to the
neutral pion field $\phi_{\pi}$, $2m_{q}P_{\pi}=-f_{\pi}m_{\pi}^{2}\phi_{\pi}$,
where $f_{\pi}$ is the pion decay constant and $m_{q}$ and $m_{\pi}$
are the quark and pion mass respectively. In the chiral limit with
zero quark mass we have $m_{\pi}=0$ indicating pions as Goldstone
bosons. 

The chiral magnetic effect (CME) is an effect closely related to the
chiral anomaly \cite{Vilenkin:1980fu,Kharzeev:2007jp,Fukushima:2008xe,Kharzeev:2015znc}.
It is about the generation of an electric current along the magnetic
field resulting from an imbalance of the population of chiral fermions.
Another accompanying effect is the vortical effect in which an electric
current is induced by the vorticity in a system of charged particles
\cite{Vilenkin:1978hb,Erdmenger:2008rm,Banerjee:2008th}. For chiral
fermions it is called the chiral vortical effect (CVE) \cite{Son:2009tf,Kharzeev:2010gr,Gao:2012ix}.
It has been demonstrated that the electric current from CME and CVE
must coexist in order to guarantee the second law of thermodynamics
in a chiral fluid \cite{Son:2009tf,Pu:2010as}. The CVE can be regarded
as a quantum effect in hydrodynamics related to chiral anomaly. 

The CME, CVE and other related effects such as chiral magnetic wave
\cite{Kharzeev:2010gd,Burnier:2011bf} have been extensively studied
in the quark-gluon plasma produced in high-energy heavy-ion collisions
in which very strong magnetic fields \cite{Kharzeev:2007jp,Skokov:2009qp,Voronyuk:2011jd,Deng:2012pc,Bloczynski:2012en,McLerran:2013hla,Gursoy:2014aka,Roy:2015coa,Tuchin:2014iua,Li:2016tel}
and huge global angular momenta \cite{Liang:2004xn,Becattini:2007sr,Betz:2007kg,Gao:2007bc}
are produced in non-central collisions. The charge separation effect
observed in STAR \cite{Abelev:2009ac,Abelev:2009ad} and ALICE \cite{Abelev:2012pa}
experiments are consistent to the CME prediction. But there were debates
that the charge separation might come from cluster particle correlations
together with a reasonable range of cluster anisotropy in nonperipheral
collisions \cite{Wang:2009kd}, so it is not conclusive that the charge
separation effect would be the evidence of the CME. The charge asymmetry
dependence of pion elliptic flow was observed in heavy-ion collisions
by STAR and is considered as the possible consequences of the chiral
magnetic wave \cite{Adamczyk:2015eqo}. The CME has recently been
confirmed to exist in materials such as Dirac and Weyl semi-metals
\cite{Son:2012bg,Basar:2013iaa,Li:2014bha}. Recently STAR collaboration
has measured nonvanishing hyperon polarization in the beam energy
scan program \cite{Lisa:2016}. This is a piece of evidence for local
polarization effect from vorticity in collisions at lower energy and
was first predicted in Ref. \cite{Liang:2004ph}. 

Quantum kinetic theory in terms of Wigner function \cite{Heinz:1983nx,Elze:1986qd,Vasak:1987um,Zhuang:1995pd}
is a useful tool to study the CME, CVE and other related effects \cite{Gao:2012ix,Chen:2012ca,Gao:2015zka,Fang:2016vpj}.
The axial vector component of the Wigner function for massless fermions
can be generalized to massive fermions and gives their phase-space
density of the spin vector. The spin vector arises from nonzero fermion
mass \cite{Chen:2013iga}. Therefore one can calculate the polarization
of massive fermions from the axial vector component \cite{Fang:2016vpj}.
The polarization density is found to be proportional to the local
vorticity $\boldsymbol{\omega}$ as well as the magnetic field. The
polarization per particle for fermions is always smaller than that
for anti-fermions as the result of more Pauli blocking effect for
fermions than antifermions. This is consistent to the STAR's preliminary
result on the $\Lambda$ polarization \cite{Lisa:2016}. 

In this paper we give another important feature of massive fermions
from the axial vector component of the Wigner function, namely, the
thermal average of pseudoscalar quantity $P$ in Eq. (\ref{eq:anomaly-2}).
In low energy QCD, it is proportional to the pion field from PCAC
hypothesis, so we call it the pseudoscalar condensate. We will show
that such a pseudoscalar condensate is induced by anomaly and force-vorticity
coupling and depends on the fermion mass, fermion chemical potential
and temperature. It is a mass effect in a plasma of fermions: for
massless fermions the pseudoscalar is vanishing. We note that there
were many studies on pseudoscalar condensates such as pion and eta
meson condensates in nuclear and quark matter in a variety of hot
and dense environments \cite{Sawyer:1973fv,Baym:1973zk,Kharzeev:1998kz,Son:2000xc,He:2005nk}
which are different from the condensates on which we focus in this
paper. 

The paper is organized as follows. In Section \ref{sec:wig-massless},
we summarize the properties of the Wigner function for massless fermions
in electromagnetic fields. In Section \ref{sec:wigner-mass} we present
the equation for pseudoscalar and axial vector component of the Wigner
function for massive fermions, the Wigner function counterpart of
Eq. (\ref{eq:anomaly-2}). In Section \ref{sec:lead-order}, we analyze
the axial vector component at leading order from which the polarization
vector can be obtained. We also derive the polarization vector of
a fermion in the lab frame with its 3-momentum $\mathbf{p}$ in the
fluid cell's comoving frame and the polarization vector $\mathbf{n}$
in the particle's rest frame. This is useful to connect the experimental
observable to the theoretical prediction. In Section \ref{sec:pseudoscalar}
we derive nonconservation of chiral current with anomaly and fermion
mass by taking space-time divergence of the chiral current derived
from the axial vector component. We also calculate condensates of
neutral pions and eta mesons. Then we derive the pseudoscalar condensate
induced by anomaly and vorticity for massive fermions. The summary
is made in the last section. 

We adopt the same sign conventions for fermion charge $Q$ as in Refs.
\cite{Vasak:1987um,Gao:2012ix,Chen:2012ca,Gao:2015zka}, and the same
sign convention for the axial vector component $\mathscr{A}^{\mu}\sim\left\langle \bar{\psi}\gamma^{\mu}\gamma^{5}\psi\right\rangle $
as in Refs. \cite{Gao:2012ix,Chen:2012ca,Gao:2015zka} but different
sign convention from Refs. \cite{Vasak:1987um}.

\section{Wigner function for massless fermions in electromagnetic fields}

\label{sec:wig-massless}The gauge invariant Wigner function is the
quantum mechanical analogue of a classical phase-space distribution.
In a background electromagnetic field, the Wigner function $W_{\alpha\beta}(x,p)$
is defined by 
\begin{equation}
W_{\alpha\beta}(x,p)=\int\frac{d^{4}y}{(2\pi)^{4}}e^{-ip\cdot y}\left\langle \bar{\psi}_{\beta}(x+\frac{1}{2}y)\mathrm{P}U(G,x+\frac{1}{2}y,x-\frac{1}{2}y)\psi_{\alpha}(x-\frac{1}{2}y)\right\rangle ,\label{eq:wigner-def-1}
\end{equation}
where $\psi_{\alpha}$ and $\bar{\psi}_{\beta}$ are fermionic quantum
fields with Dirac indices $\alpha$ and $\beta$, $\langle\hat{O}\rangle$
denotes the grand canonical ensemble average of normal ordered operator,
$x=(x_{0},\mathbf{x})$ and $p=(p_{0},\mathbf{p})$ are time-space
and energy-momentum 4-vectors respectively, and the gauge link $\mathrm{P}U(G,x_{1},x_{2})$
is to ensure the gauge invariance of the Wigner function where $G^{\mu}$
is the gauge potential of the classical electromagnetic field and
P denotes the path-ordered product. The $4\times4$ matrix $W_{\alpha\beta}(x,p)$
can be decomposed by 16 independent generators of Clifford algebra,
namely,$1,\gamma^{5},\gamma^{\mu},\gamma^{5}\gamma^{\mu},\sigma^{\mu\nu}\equiv\frac{i}{2}[\gamma^{\mu},\gamma^{\nu}]$,
into the scalar, pseudoscalar, vector, axial vector and tensor components,
respectively. From the Dirac equation for the fermionic field, one
can derive the equation for $W_{\alpha\beta}(x,p)$ which leads to
a set of coupled equations for all components. Finding a general solution
to the Wigner function is very difficult. However it is much simplified
for massless fermions for which the set of equations for the vector
and axial vector components are decoupled from the rest components.
Assuming that the electromagnetic field is homogeneous and weak and
is in the same order as the space-time derivative, one can solve the
vector and axial vector components perturbatively. To the linear order
in the field strength and vorticity, one can obtain the vector and
axial vector components \cite{Gao:2012ix}, which give the charge
current $j^{\mu}$ and chiral (axial) charge current $j_{5}^{\mu}$
by integration over four-momenta: $j^{\mu}=nu^{\mu}+\xi\omega^{\mu}+\xi_{B}B^{\mu}$
and $j_{5}^{\mu}=nu^{\mu}+\xi_{5}\omega^{\mu}+\xi_{5B}B^{\mu}$. Here
$\omega^{\mu}=\frac{1}{2}\epsilon^{\mu\nu\rho\sigma}u_{\nu}\partial_{\rho}u_{\sigma}$
is the vorticity vector, $B^{\mu}=\frac{1}{2}\epsilon^{\mu\nu\rho\sigma}u_{\nu}F_{\rho\sigma}$
is the magnetic field four-vector with the fluid four-velocity $u_{\nu}$,
$n$ and $n_{5}$ are charge and chiral (axial) charge density respectively.
In the charge current $j^{\mu}$ one obtains the CME and CVE coefficients
$\xi=\mu\mu_{5}/\pi^{2}$ and $\xi_{B}=Q\mu_{5}/(2\pi^{2})$ respectively,
where $\mu$ and $\mu_{5}$ are chemical potentials for the charge
and chiral (axial) charge respectively. One can also derive the coefficients
of vorticty and magnetic field in $j_{5}^{\mu}$: $\xi_{5}=T^{2}/6+(\mu^{2}+\mu_{5}^{2})/(2\pi^{2})$
and $\xi_{5B}=Q\mu/(2\pi^{2})$. The conservation and anomalous nonconservation
laws for the charge and chiral (axial) charge current respectively
can be verified, $\partial_{\mu}j^{\mu}=0$ and $\partial_{\mu}j_{5}^{\mu}=-[Q^{2}/(8\pi^{2})]F_{\mu\nu}\tilde{F}^{\mu\nu}$.
The covariant chiral kinetic equation which is related to the Berry
phase in 4-dimension can also be derived from the first order solution
to the vector and axial vector components of the Wigner function \cite{Chen:2012ca}.

\section{Equation for pseudoscalar and axial vector component of Wigner function}

\label{sec:wigner-mass}In this section we look at the equation for
pseudoscalar and axial vector component of the Wigner function. From
Dirac equation for the fermionic field, one can derive the equation
for the Wigner function in (\ref{eq:wigner-def-1}) in a constant
electromagnetic field, 
\begin{equation}
\left[\gamma_{\mu}\left(p^{\mu}+i\hbar\frac{1}{2}\nabla^{\mu}\right)-m\right]W(x,p)=0,\label{eq:wigner-dirac}
\end{equation}
where the phase-space derivative is defined by $\nabla^{\mu}=\partial_{x}^{\mu}-QF^{\mu\nu}\partial_{p,\nu}$
and we suppressed Dirac indices of the Wigner function. The Wigner
function as a $4\times4$ matrix in Dirac space can be decomposed
into the scalar, pseudoscalar, vector, axial vector and tensor components
as 
\begin{equation}
W=\frac{1}{4}\left[\mathscr{F}+i\gamma^{5}\mathscr{P}+\gamma^{\mu}\mathscr{V}_{\mu}+\gamma^{5}\gamma^{\mu}\mathscr{A}_{\mu}+\frac{1}{2}\sigma^{\mu\nu}\mathscr{S}_{\mu\nu}\right].\label{eq:wigner-decomp}
\end{equation}
The components in the decomposition (\ref{eq:wigner-decomp}) can
be obtained by projection of corresponding Dirac matrices on the Wigner
function and taking traces. Eq. (\ref{eq:wigner-dirac}) for the Wigner
function can be converted to a set of coupled equations for all components.
There is an equation that relates the pseudoscalar to the axial vector
component which is of special interest, 
\begin{eqnarray}
\hbar\nabla^{\mu}\mathscr{A}_{\mu} & = & -2m\mathscr{P}.\label{eq:real}
\end{eqnarray}
An interesting observation of the above equation is that the pseudoscalar
component $\mathscr{P}$ is of quantum origin since it is proportional
to the Planck constant $\hbar$. We note that Eq. (\ref{eq:real})
is nothing but the Wigner function counterpart of Eq. (\ref{eq:anomaly-2}).
We will show that the integration of Eq. (\ref{eq:real}) over 4-momentum
gives Eq. (\ref{eq:anomaly-2}).

\section{Axial vector component at leading order and polarization vector}

\label{sec:lead-order}At leading (zero-th) order of electromagnetic
interaction, the gauge link in the Wigner function in Eq. (\ref{eq:wigner-def-1})
can be set to 1, we denote the Wigner function at this order as $W_{(0)}$.
We can expand fermionic fields in momentum space with creation and
destruction operators, which we insert into Eq. (\ref{eq:wigner-def-1}).
After taking ensemble average of normal ordered operators, we obtain
\begin{eqnarray}
W_{(0)}(x,p) & = & \frac{1}{(2\pi)^{3}}\delta(p^{2}-m^{2})\bigg\{\theta(p^{0})\sum_{s}f_{\mathrm{FD}}(E_{p}-\mu_{s})u(\mathbf{p},s)\bar{u}(\mathbf{p},s)\nonumber \\
 &  & -\theta(-p^{0})\sum_{s}f_{\mathrm{FD}}(E_{p}+\mu_{s})v(-\mathbf{p},s)\bar{v}(-\mathbf{p},s)\bigg\},\label{eq:wigner-spinor}
\end{eqnarray}
where $u(\mathbf{p},s)$ and $v(-\mathbf{p},s)$ are Dirac spinors
of positive and negative energy respectively, $s=\pm$ denote the
spin state parallel or anti-parallel to the spin quantization direction
$\mathbf{n}$ in the rest frame of the particle. We have also used
$\left\langle a^{\dagger}(\mathbf{p},s)a(\mathbf{p},s)\right\rangle =f_{\mathrm{FD}}(E_{p}-\mu_{s})$
and $\left\langle b^{\dagger}(-\mathbf{p},s)b(-\mathbf{p},s)\right\rangle =f_{\mathrm{FD}}(E_{p}+\mu_{s})$
with the Fermi-Dirac distribution defined by $f_{\mathrm{FD}}(x)=1/(e^{\beta x}+1)$
($\beta\equiv1/T$, $T$ is temperature) and $\mu_{s}$ is the chemical
potential for fermions in the spin state $s$. 

The axial vector component at the leading order is given by 
\begin{eqnarray}
\mathscr{A}_{(0)}^{\mu} & = & \mathrm{Tr}[\gamma^{\mu}\gamma^{5}W_{(0)}]\nonumber \\
 & = & m\left[\theta(p_{0})n^{\mu}(\mathbf{p},\mathbf{n})-\theta(-p_{0})n^{\mu}(-\mathbf{p},-\mathbf{n})\right]\delta(p^{2}-m^{2})A,\label{eq:a0-1}
\end{eqnarray}
where $A$ is defined by 
\begin{equation}
A\equiv\frac{2}{(2\pi)^{3}}\sum_{s}s\left[\theta(p^{0})f_{\mathrm{FD}}(p_{0}-\mu_{s})+\theta(-p^{0})f_{\mathrm{FD}}(-p_{0}+\mu_{s})\right],\label{eq:axial-const}
\end{equation}
and we have used $\bar{u}(\mathbf{p},s)\gamma^{\mu}\gamma^{5}u(\mathbf{p},s)=2msn^{\mu}(\mathbf{p},\mathbf{n})$
and $\bar{v}(-\mathbf{p},s)\gamma^{\mu}\gamma^{5}v(-\mathbf{p},s)=2msn^{\mu}(-\mathbf{p},-\mathbf{n})$
with $n^{\mu}(\mathbf{p},\mathbf{n})$ given by 
\begin{eqnarray}
n^{\mu}(\mathbf{p},\mathbf{n}) & = & \Lambda_{\;\nu}^{\mu}(-\mathbf{v}_{p})n^{\nu}(\mathbf{0},\mathbf{n})=\left(\frac{\mathbf{n}\cdot\mathbf{p}}{m},\mathbf{n}+\frac{(\mathbf{n}\cdot\mathbf{p})\mathbf{p}}{m(m+E_{p})}\right).\label{eq:polar-cmoving}
\end{eqnarray}
Here $\Lambda_{\;\nu}^{\mu}(-\mathbf{v}_{p})$ is the Lorentz transformation
for $\mathbf{v}_{p}=\mathbf{p}/E_{p}$ and $n^{\nu}(\mathbf{0},\mathbf{n})=(0,\mathbf{n})$
is the 4-vector of the spin quantization direction in the rest frame
of the fermion. One can check that $n^{\mu}(\mathbf{p},\mathbf{n})$
satisfies $n^{2}=-1$ and $n\cdot p=0$, so it behaves like a spin
4-vector up to a factor of 1/2. For Pauli spinors $\chi_{s}$ and
$\chi_{s^{\prime}}$ in $u(\mathbf{p},s)$ and $v(-\mathbf{p},s^{\prime})$
respectively, we have $\chi_{s}^{\dagger}\boldsymbol{\sigma}\chi_{s}=s\mathbf{n}$
and $\chi_{s^{\prime}}^{\dagger}\boldsymbol{\sigma}\chi_{s^{\prime}}=-s^{\prime}\mathbf{n}$.
We can take the massless limit by setting $\mathbf{n}=\hat{\mathbf{p}}$,
then we have $mn^{\mu}(\mathbf{p},\mathbf{n})\rightarrow(|\mathbf{p}|,\mathbf{p})$
and $mn^{\mu}(-\mathbf{p},-\mathbf{n})\rightarrow(|\mathbf{p}|,-\mathbf{p})$.
This way we can recover the previous result of the axial vector component
for massless fermions \cite{Gao:2012ix,Chen:2012ca} where $s=\pm$
denote the right-handed and left-handed fermions. 

We implied that $n^{\mu}(\mathbf{p},\mathbf{n})$ in (\ref{eq:polar-cmoving})
is the form in the comoving or local rest frame of a fluid cell. We
can boost all quantities in Eq. (\ref{eq:a0-1}) to the lab frame
in which the fluid cell is moving with a 4-velocity $u^{\alpha}=\gamma(1,\mathbf{v})$
where $\gamma=1/\sqrt{1-v^{2}}$ is the Lorentz factor. We boost $n^{\mu}(\mathbf{p},\mathbf{n})$
and $n^{\mu}(-\mathbf{p},-\mathbf{n})$ to the lab frame as 
\begin{eqnarray}
n_{\mathrm{lab}}^{\prime\alpha}(\mathbf{p},\mathbf{n}) & = & \Lambda_{\;\beta}^{\alpha}(-\mathbf{v})n^{\beta}(\mathbf{p},\mathbf{n}),\nonumber \\
n_{\mathrm{lab}}^{\prime0}(\mathbf{p},\mathbf{n}) & = & \gamma\left[\frac{\mathbf{n}\cdot\mathbf{p}}{m}+\mathbf{v}\cdot\mathbf{n}+\frac{(\mathbf{n}\cdot\mathbf{p})(\mathbf{v}\cdot\mathbf{p})}{m(m+E_{p})}\right],\nonumber \\
\mathbf{n}_{\mathrm{lab}}^{\prime}(\mathbf{p},\mathbf{n}) & = & \mathbf{n}+\frac{(\mathbf{n}\cdot\mathbf{p})\mathbf{p}}{m(m+E_{p})}+\frac{\gamma-1}{v^{2}}\mathbf{v}\left[\mathbf{n}\cdot\mathbf{v}+\frac{(\mathbf{n}\cdot\mathbf{p})(\mathbf{p}\cdot\mathbf{v})}{m(m+E_{p})}\right]+\frac{\gamma}{m}\mathbf{v}(\mathbf{n}\cdot\mathbf{p}).\label{eq:polar-cm}
\end{eqnarray}
Note that $\mathbf{n}$ is the spin quantization direction in the
fermion's rest frame and $\mathbf{p}$ is the fermion's momentum in
the local rest frame of the fluid cell. The fermion momentum in the
lab frame is $p^{\prime\alpha}=\Lambda_{\;\beta}^{\alpha}(-\mathbf{v})p^{\beta}$
with $p^{\beta}=(E_{p},\mathbf{p})$ or explicitly 
\begin{align}
p^{\prime0} & =\gamma[E_{p}+(\mathbf{p}\cdot\mathbf{v})],\nonumber \\
\mathbf{p}^{\prime} & =\mathbf{p}+\frac{\gamma-1}{v^{2}}\mathbf{v}(\mathbf{p}\cdot\mathbf{v})+\gamma\mathbf{v}E_{p}.\label{eq:p-prime}
\end{align}
One can check $p^{\prime}\cdot u=p^{0}=E_{p}$ and $n^{\prime}\cdot p^{\prime}=0$
in the lab frame. So Eq. (\ref{eq:a0-1}) and (\ref{eq:axial-const})
can be written in the lab frame by making the replacement $n^{\alpha}(\mathbf{p},\mathbf{n})\rightarrow n_{\mathrm{lab}}^{\prime\alpha}(\mathbf{p},\mathbf{n})$,
also we have $p_{0}=p^{\prime}\cdot u$ and $p^{\prime2}=p^{2}$ in
the two formula. 

We note that Eq. (\ref{eq:polar-cm}) is the polarization vector in
the lab frame (the fluid cell moves with a velocity $\mathbf{v}$)
of a fermion with the 3-momentum $\mathbf{p}$ in the fluid cell's
comoving frame and the polarization vector $\mathbf{n}$ in the particle's
rest frame. After taking integration of $n_{\mathrm{lab}}^{\prime\alpha}(\mathbf{p},\mathbf{n})$
over $\mathbf{p}$ which follows the Fermi-Dirac distribution, one
can obtain the thermal average of the polarization vector in the lab
frame. The $\Lambda$ polarization can be measured in experiments
by its decay to proton and pion in its rest frame. It is observed
that $\Lambda$ is polarized along the global angular momentum in
the beam energy scan program at RHIC \cite{Lisa:2016}. From Eq. (\ref{eq:polar-cm})
one can calculate the $\Lambda$ polarization along a fixed direction
in the lab frame, e.g. the direction of the global angular momentum,
and compare with data. 

In this paper, we assume that the chemical potential does not depend
on the spin state $\mu_{s}=\mu$ for $s=\pm$, so $A=0$ and then
$\mathscr{A}_{(0)}^{\mu}$ is vanishing. The nonvanishing contribution
comes from the axial vector component at the next-to-leading or first
order even with spin-independent chemical potentials.

\section{Chiral current nonconservation law and pseudoscalar condensation}

\label{sec:pseudoscalar}In this section we will derive nonconservation
law of chiral current of massive fermions by taking space-time divergence
of the chiral current derived from the axial vector component. Then
we can derive the pseudoscalar condensate induced by anomaly and vorticity. 

\begin{figure}
\noindent \includegraphics[scale=0.7]{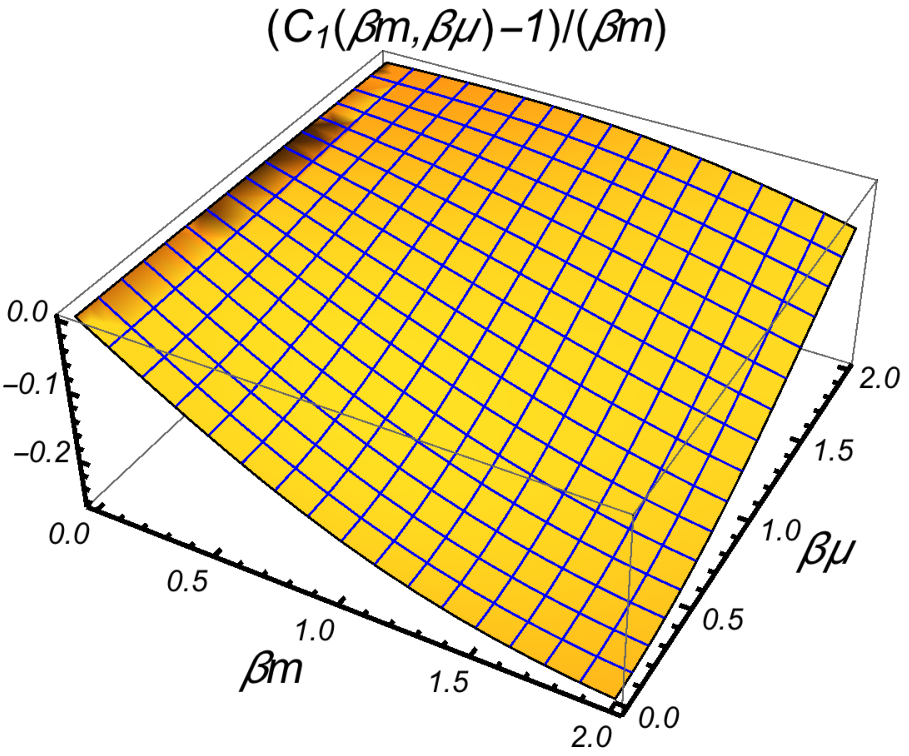} \includegraphics[scale=0.7]{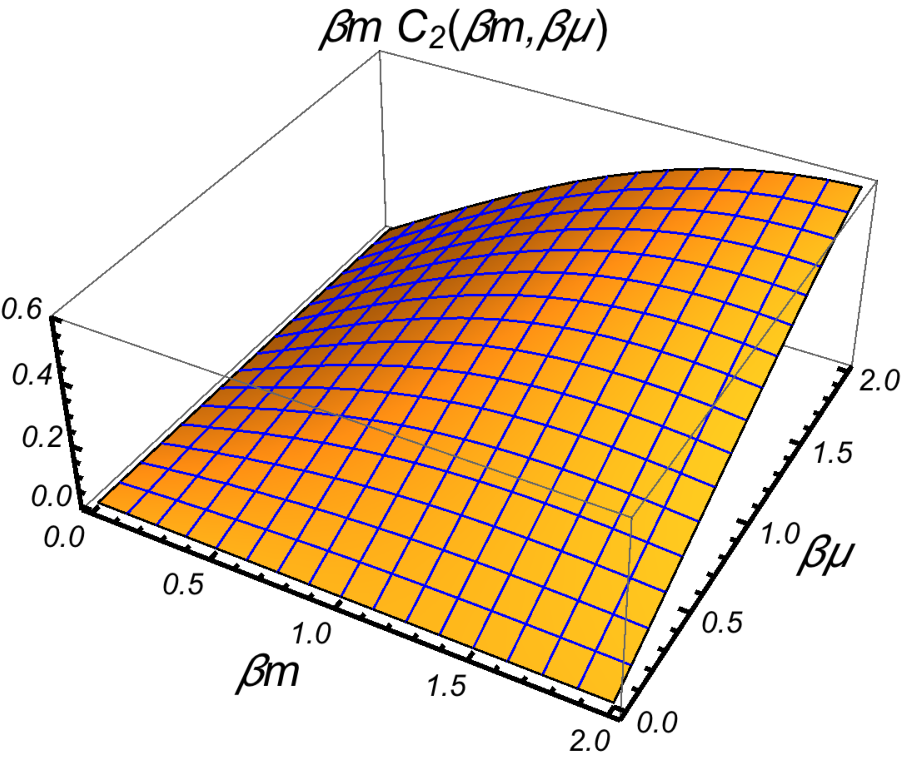}

\caption{\label{fig:c1_c2} $[C_{1}(\beta m,\beta\mu)-1]/(\beta m)$ and $(\beta m)C_{2}(\beta m,\beta\mu)$
as functions of $\beta m$ and $\beta\mu$. They are all proportional
to $m$ in the chiral limit. }

\end{figure}

As we have discussed in the last section that the contribution should
come from the axial vector component at the next-to-leading or first
order. We use the following form for the axial vector component for
massive fermions by generalizing the solution for massless fermions
\cite{Gao:2012ix,Chen:2012ca,Gao:2015zka,Fang:2016vpj}, 
\begin{eqnarray}
\mathscr{A}_{(1)}^{\alpha}(x,p) & = & -\frac{1}{2}\hbar\beta\tilde{\Omega}^{\alpha\sigma}p_{\sigma}\frac{dV}{d(\beta p_{0})}\delta(p^{2}-m^{2})-Q\hbar\tilde{F}^{\alpha\lambda}p_{\lambda}V\frac{\delta(p^{2}-m^{2})}{p^{2}-m^{2}},\label{eq:a1}
\end{eqnarray}
where $p_{0}\equiv u\cdot p$, $\tilde{\Omega}^{\rho\sigma}=\frac{1}{2}\epsilon^{\rho\sigma\mu\nu}\partial_{\mu}u_{\nu}$,
and $V$ is associated with the vector component and given by 
\begin{equation}
V\equiv\frac{4}{(2\pi)^{3}}\left[\theta(p^{0})f_{\mathrm{FD}}(p_{0}-\mu)+\theta(-p^{0})f_{\mathrm{FD}}(-p_{0}+\mu)\right].\label{eq:vector-const}
\end{equation}
where we have taken $\mu_{s}=\mu$ for $s=\pm$. Also we have assumed
in Eq. (\ref{eq:a1}) that $\beta=1/T$ is a constant. The chiral
current can be obtained by integrating over 4-momentum of the axial
vector component, $j_{5}^{\mu}=\int d^{4}p\mathscr{A}_{(1)}^{\mu}(x,p)$,
whose space-time divergence is given by 
\begin{eqnarray}
\partial_{\mu}j_{5}^{\mu} & = & \int d^{4}p\partial_{\mu}\mathscr{A}_{(1)}^{\mu}\nonumber \\
 & = & -\frac{1}{2}\hbar\beta\tilde{\Omega}^{\mu\sigma}\int d^{4}pp_{\sigma}\partial_{\mu}\left[\frac{dV}{d(\beta p_{0})}\right]\delta(p^{2}-m^{2})\nonumber \\
 &  & -\hbar Q\tilde{F}^{\mu\lambda}\int d^{4}pp_{\lambda}(\partial_{\mu}V)\frac{\delta(p^{2}-m^{2})}{p^{2}-m^{2}}.\label{eq:d-j5}
\end{eqnarray}
where we have assumed that $\tilde{F}^{\mu\lambda}$ does not depend
on space-time, and used $u_{\sigma}\partial_{\mu}\tilde{\Omega}^{\mu\sigma}=0$
due to static equilibrium conditions \cite{Gao:2012ix}. When taking
space-time divergence, we neglect the space-time derivative of $\theta(p_{0})$
and $\theta(-p_{0})$ in $V$ which would be vanishing when contacting
the mass-shell condition. The two terms in the last equality of Eq.
(\ref{eq:d-j5}) are evaluated in Appendix \ref{sec:derivation-d-j5}
and can be grouped into a $E\cdot B$ and $E\cdot\omega$ term, 
\begin{eqnarray}
\partial_{\mu}j_{5}^{\mu} & = & \frac{1}{2}\hbar Q^{2}\beta^{2}(E\cdot B)\int d^{4}pV_{\beta\mu,\beta p_{0}}^{\prime\prime}\delta(p^{2}-m^{2})\nonumber \\
 &  & +\frac{1}{2}\hbar\beta Q(E\cdot\omega)\int d^{4}p\left[\beta p_{0}V_{\beta p_{0},\beta\mu}^{\prime\prime}-\beta\frac{2\bar{p}^{2}}{3p_{0}}V_{\beta p_{0},\beta p_{0}}^{\prime\prime}+\frac{2\bar{p}^{2}}{3p_{0}^{2}}V_{\beta p_{0}}^{\prime}\right]\delta(p^{2}-m^{2}),\label{eq:i23-2}
\end{eqnarray}
where $\bar{p}^{\alpha}\equiv p^{\alpha}-(p\cdot u)u^{\alpha}$. The
integrals in Eq. (\ref{eq:i23-2}) can be finally simplified into
following forms 
\begin{eqnarray}
\partial_{\mu}j_{5}^{\mu} & = & -\frac{1}{2\pi^{2}}\hbar Q^{2}(E\cdot B)C_{1}(\beta m,\beta\mu)-\frac{m^{2}}{2\pi^{2}}\hbar Q\beta(E\cdot\omega)C_{2}(\beta m,\beta\mu),\label{eq:i23-final}
\end{eqnarray}
where the dimensionless functions $C_{1}(\beta m,\beta\mu)$ and $C_{2}(\beta m,\beta\mu)$
are defined by 
\begin{eqnarray}
C_{1}(\beta m,\beta\mu) & = & \int_{0}^{\infty}dx\left[\frac{e^{\sqrt{x^{2}+(\beta m)^{2}}-\beta\mu}}{(e^{\sqrt{x^{2}+(\beta m)^{2}}-\beta\mu}+1)^{2}}+\frac{e^{\sqrt{x^{2}+(\beta m)^{2}}+\beta\mu}}{(e^{\sqrt{x^{2}+(\beta m)^{2}}+\beta\mu}+1)^{2}}\right],\nonumber \\
C_{2}(\beta m,\beta\mu) & = & \int_{0}^{\infty}dx\frac{1}{\sqrt{x^{2}+(\beta m)^{2}}}\left[\frac{e^{\sqrt{x^{2}+(\beta m)^{2}}-\beta\mu}}{(e^{\sqrt{x^{2}+(\beta m)^{2}}-\beta\mu}+1)^{2}}-\frac{e^{\sqrt{x^{2}+(\beta m)^{2}}+\beta\mu}}{(e^{\sqrt{x^{2}+(\beta m)^{2}}+\beta\mu}+1)^{2}}\right].\label{eq:c1c2}
\end{eqnarray}
We plot the functions $C_{1}(\beta m,\beta\mu)-1$ and $C_{2}(\beta m,\beta\mu)$
in Fig. \ref{fig:c1_c2}. One can verify asymptotic values $C_{1}(\beta m,\beta\mu)-1\sim O[(\beta m)^{2}]$
and $C_{2}(\beta m,\beta\mu)\sim O(1)$ for $\beta m\rightarrow0$,
see Eq. (\ref{eq:asymptotic-c1c2}) in Appendix \ref{sec:c1c2}. 

On the other hand, taking an integration over 4-momentum of Eq. (\ref{eq:real})
leads to 
\begin{eqnarray}
-2mP & = & \hbar\partial_{\mu}j_{5}^{\mu}+\frac{1}{2\pi^{2}}\hbar^{2}Q^{2}(E\cdot B),\label{eq:pseudoscalar}
\end{eqnarray}
where we have used $P(x)=\int d^{4}p\mathscr{P}(x,p)$ and the inetgral
\begin{equation}
QF_{\mu\nu}\int d^{4}p\partial_{p}^{\nu}\mathscr{A}^{\mu}(x,p)=-\frac{1}{2\pi^{2}}\hbar Q^{2}(E\cdot B),\label{eq:p-derivative}
\end{equation}
whose derivation is given in Appendix \ref{sec:p-derivative}. Inserting
Eq. (\ref{eq:i23-final}) into Eq. (\ref{eq:pseudoscalar}) we obtain
the pseudoscalar condensate 
\begin{eqnarray}
P & = & \frac{1}{4\pi^{2}}\hbar^{2}Q^{2}(E\cdot B)\frac{1}{m}[C_{1}(\beta m,\beta\mu)-1]+\frac{1}{4\pi^{2}}\hbar^{2}Q(E\cdot\omega)\beta mC_{2}(\beta m,\beta\mu).\label{eq:pseudoscalar-1}
\end{eqnarray}
From the small mass behavior $C_{1}(\beta m,\beta\mu)-1\sim(\beta m)^{2}$
and $C_{2}(\beta m,\beta\mu)\sim O(1)$, the pseudoscalar is proportional
to the fermion mass. This feature is similar to the PCAC hypothesis
where the pseudoscalar is proportional to $m_{\pi}^{2}/m_{q}$. We
see in $C_{2}(\beta m,\beta\mu)$ that there is a sign difference
between fermion and antifermion terms, so the $QE\cdot\omega$ term
can also be regarded as the \textit{force-vorticity} coupling term.
Furthermore the $E\cdot\omega$ term is vanishing for $\beta\mu=0$. 

The above discussions are about single fermion species. Let us consider
a quark plasma with two quark flavors $u$ and $d$. For $u$ or $d$
quarks, Eqs. (\ref{eq:i23-final},\ref{eq:pseudoscalar}) become 
\begin{eqnarray}
\partial_{\mu}j_{5,a}^{\mu} & = & -\frac{N_{C}}{2\pi^{2}}\hbar Q_{a}^{2}(E\cdot B)C_{1}(\beta m_{a},\beta\mu_{a})-\frac{N_{C}m_{a}^{2}}{2\pi^{2}}\hbar Q_{a}\beta(E\cdot\omega)C_{2}(\beta m_{a},\beta\mu_{a}),\nonumber \\
-2m_{a}P_{a} & = & \hbar\partial_{\mu}j_{5,a}^{\mu}+\frac{N_{C}}{2\pi^{2}}\hbar^{2}Q_{a}^{2}(E\cdot B),\label{eq:pion-anomaly}
\end{eqnarray}
where $N_{C}=3$ is the number of color, $a=u,d$, and $P_{a}=-i\bar{a}\gamma_{5}a$,
$j_{5,a}^{\mu}=\bar{a}\gamma^{\mu}\gamma_{5}a$. We know from the
introduction that the pseudoscalar and chiral current for neutral
pions are given by
\begin{eqnarray}
P_{\pi} & = & -i\bar{\psi}\gamma_{5}(\sigma_{3}/2)\psi=\frac{1}{2}(P_{u}-P_{d}),\nonumber \\
j_{5,\pi}^{\mu} & = & \bar{\psi}\gamma^{\mu}\gamma_{5}(\sigma_{3}/2)\psi=\frac{1}{2}(j_{5,u}^{\mu}-j_{5,d}^{\mu}).\label{eq:pion-p-j}
\end{eqnarray}
In the first case we assume that the masses and chemical potentials
of $u$ and $d$ quarks are the same, $m_{u}=m_{d}=m_{q}$ and $\mu_{u}=\mu_{d}=\mu_{q}$.
Then the pion condensate is 
\begin{eqnarray}
P_{\pi} & = & \frac{N_{C}}{8\pi^{2}}\hbar^{2}(Q_{u}^{2}-Q_{d}^{2})(E\cdot B)\frac{1}{m_{q}}[C_{1}(\beta m_{q},\beta\mu_{q})-1]\nonumber \\
 &  & +\frac{N_{C}}{8\pi^{2}}\hbar^{2}(Q_{u}-Q_{d})(E\cdot\omega)\beta m_{q}C_{2}(\beta m_{q},\beta\mu_{q})\nonumber \\
 & = & \frac{N_{C}}{24\pi^{2}}\hbar^{2}Q_{e}^{2}(E\cdot B)\frac{1}{m_{q}}[C_{1}(\beta m_{q},\beta\mu_{q})-1]+\frac{N_{C}}{8\pi^{2}}\hbar^{2}Q_{e}(E\cdot\omega)\beta m_{q}C_{2}(\beta m_{q},\beta\mu_{q})\label{eq:pion-condensate}
\end{eqnarray}
In the second case we assume that the masses of $u$ and $d$ quarks
are the same, $m_{u}=m_{d}=m_{q}$, but the chemical potentials are
different, the pion condensate is 
\begin{eqnarray}
P_{\pi} & = & \frac{N_{C}}{8\pi^{2}}\hbar^{2}(E\cdot B)\frac{1}{m_{q}}\left[Q_{u}^{2}C_{1}(\beta m_{q},\beta\mu_{u})-Q_{d}^{2}C_{1}(\beta m_{q},\beta\mu_{d})-(Q_{u}^{2}-Q_{d}^{2})\right]\nonumber \\
 &  & +\frac{N_{C}}{8\pi^{2}}\hbar^{2}(E\cdot\omega)\beta m_{q}\left[Q_{u}C_{2}(\beta m_{q},\beta\mu_{u})-Q_{d}C_{2}(\beta m_{q},\beta\mu_{d})\right]\label{eq:pion-cond-general}
\end{eqnarray}
In vacuum at zero temperature and chemical potentials, both functions
$C_{1}$ and $C_{2}$ are vanishing and we obtain from Eqs. (\ref{eq:pion-condensate},\ref{eq:pion-cond-general})
\begin{equation}
P_{\pi}^{\mathrm{vac}}=-\frac{\hbar^{2}}{8\pi^{2}m_{q}}Q_{e}^{2}(E\cdot B),\label{eq:pion-cond-vac}
\end{equation}
which is consistent with the result derived in the NJL model and chiral
perturbation theory \cite{Cao:2015cka}. We see that the pion condensate
in vacuum comes out quite natural as a mass effect of charged fermions
and is not subject to any additional constraint such as a critical
value for |$\mathbf{E}\cdot\mathbf{B}$| in \cite{Cao:2015cka}. Recently
the chiral current and pseudoscalar condensate for massive fermions
have also been calculated in a holographic model in finite density
and magnetic field \cite{Guo:2016dnm}. 

We can also generalize to quark matter with three quark flavors $u$,
$d$ and $s$. So Eq. (\ref{eq:pion-anomaly}) applies to $a=u,d,s$.
Now we look at the pseudoscalar and chiral current for the flavor
octet $\eta_{8}$ and singlet $\eta_{1}$ which can be defined by
\begin{eqnarray}
P_{\eta8} & = & -i\bar{\psi}\gamma_{5}\frac{\lambda_{8}}{2}\psi=\frac{1}{2\sqrt{3}}(P_{u}+P_{d}-2P_{s}),\nonumber \\
P_{\eta1} & = & -i\bar{\psi}\gamma_{5}\frac{1}{\sqrt{3}}\boldsymbol{1}\psi=\frac{1}{\sqrt{3}}(P_{u}+P_{d}+P_{s}),\nonumber \\
j_{5,\eta8}^{\mu} & = & \bar{\psi}\gamma^{\mu}\gamma_{5}\frac{\lambda_{8}}{2}\psi=\frac{1}{2\sqrt{3}}\left(j_{5,u}^{\mu}+j_{5,d}^{\mu}-2j_{5,s}^{\mu}\right),\nonumber \\
j_{5,\eta1}^{\mu} & = & \bar{\psi}\gamma^{\mu}\gamma_{5}\frac{1}{\sqrt{3}}\boldsymbol{1}\psi=\frac{1}{\sqrt{3}}\left(j_{5,u}^{\mu}+j_{5,d}^{\mu}+j_{5,s}^{\mu}\right).
\end{eqnarray}
As an example for the $\eta$ meson condensate, we consider the simplest
case for quark masses and chemical potentials, $m_{u}=m_{d}=m_{s}=m_{q}$
and $\mu_{u}=\mu_{d}=\mu_{s}=\mu_{q}$, which can be considered as
the case of approaching the chiral limit. In this case, we have 
\begin{eqnarray}
P_{\eta8} & = & \frac{N_{C}}{8\sqrt{3}\pi^{2}}\hbar^{2}(Q_{u}^{2}+Q_{d}^{2}-2Q_{s}^{2})(E\cdot B)\frac{1}{m_{q}}[C_{1}(\beta m_{q},\beta\mu_{q})-1]\nonumber \\
 &  & +\frac{N_{C}}{8\sqrt{3}\pi^{2}}\hbar^{2}(Q_{u}+Q_{d}-2Q_{s})(E\cdot\omega)\beta m_{q}C_{2}(\beta m_{q},\beta\mu_{q})\nonumber \\
 & = & \frac{N_{C}}{24\sqrt{3}\pi^{2}}\hbar^{2}Q_{e}^{2}(E\cdot B)\frac{1}{m_{q}}[C_{1}(\beta m_{q},\beta\mu_{q})-1]+\frac{N_{C}}{8\sqrt{3}\pi^{2}}\hbar^{2}Q_{e}(E\cdot\omega)\beta m_{q}C_{2}(\beta m_{q},\beta\mu_{q})\nonumber \\
P_{\eta1} & = & \frac{N_{C}}{4\sqrt{3}\pi^{2}}\hbar^{2}(Q_{u}^{2}+Q_{d}^{2}+Q_{s}^{2})(E\cdot B)\frac{1}{m_{q}}[C_{1}(\beta m_{q},\beta\mu_{q})-1]\nonumber \\
 &  & +\frac{N_{C}}{4\sqrt{3}\pi^{2}}\hbar^{2}(Q_{u}+Q_{d}+Q_{s})(E\cdot\omega)\beta m_{q}C_{2}(\beta m_{q},\beta\mu_{q})\nonumber \\
 & = & \frac{N_{C}}{6\sqrt{3}\pi^{2}}\hbar^{2}Q_{e}^{2}(E\cdot B)\frac{1}{m_{q}}[C_{1}(\beta m_{q},\beta\mu_{q})-1].
\end{eqnarray}
The main difference between $P_{\eta8}$ and $P_{\eta1}$ in the above
is that $P_{\eta1}$ does not have $E\cdot\omega$ term due to the
cancellation of electric charges of quarks with different flavors.
Of course more realistic cases in heavy ion collisions should be like
$m_{u}=m_{d}=m_{q}\neq m_{s}$ and $\mu_{u}=\mu_{d}=\mu_{q}\neq\mu_{s}$,
for which the calculation is straightforward. There are possible observables
of $\eta$ meson condensates in heavy ion collisions \cite{Kharzeev:1998kz},
but it is beyond the scope of the current paper and will be addressed
in a future study. 

The pseudoscalar condensate for charged fermions in Eq. (\ref{eq:pseudoscalar-1})
is our main result. It is a quite general formula. Such a pseudoscalar
condensate is charge neutral and induced by anomaly and vorticity
in thermal and dense environment, which is a natural consequence of
nonconservation law of the chiral current in electromagnetic fields
and is not subject to any additional constraints. For example, the
neutral pion condensate is always in the forms of Eqs. (\ref{eq:pion-condensate},\ref{eq:pion-cond-general})
without further condition about the value of |$\mathbf{E}\cdot\mathbf{B}$|
if anomaly and vorticity are there. It is also worth mentioning a
new electric-field-vorticity coupling term in the condensate which
has not been derived in literature to our knowledge. Such pion and
eta meson condensates may have observables related to the electromagnetic
field and vorticity in heavy ion collisions. For example they may
have effects on the collective flows of neutral pions and eta mesons.
In some sense they are similar to disoriented chiral condensates \cite{Blaizot:1992at,Rajagopal:1995bc,Mohanty:2005mv}.
This is a topic that we will address in the future.

\section{Summary}

We derive the pseudoscalar condensate induced by anomaly and vorticity
from the Wigner function for massive fermions in homogeneous electromagnetic
fields. The pseudoscalar component of the Wigner function is determined
from the axial vector component by Eq. (\ref{eq:real}). Taking an
integration over 4-momentum for Eq. (\ref{eq:real}) we obtain the
anomalous nonconservation of the chiral current by the anomalous term
and a product of mass and pseudoscalar. By directly calculating the
space-time divergence of the chiral current, we can determine the
pseudoscalar condensate which has an anomalous $E\cdot B$ term and
an $E\cdot\omega$ term. The $E\cdot\omega$ term can also be regarded
as a force-vorticity coupling since there is a sign difference in
its prefactor between the fermion and antifermion sector. The force-vorticity
part of the pseudoscalar condensate is the new term. As a mass effect,
the pseudoscalar condensate is linearly proportinal to the fermion
mass when the mass is small. Such a pseudoscalar condensate is a general
feature for a fluid of massive and charged fermions in a thermal and
dense plasma with anomaly and vorticity. The neutral pion and eta
meson condensates can also be derived from generalization of the single
flavor to multi-flavor case, which depend on quark masses, quark chemical
potentials and temperature. We can reproduce the previous result of
the neutral pion condensate in vacuum induced by the anomaly, but
our result also has a force-vorticity part which has not been derived
in previous literature to our knowledge. There are possible observables
of pseudoscalar condensates related to the electromagnetic field and
vorticity in heavy ion collisions such as collective flows of neutral
pions and eta mesons. 

\textit{Acknowledgments.} 

QW thanks Xu-guang Huang and Shu Lin for helpful discussions. RHF
and QW are supported in part by the Major State Basic Research Development
Program (MSBRD) in China under the Grant No. 2015CB856902 and 2014CB845402
and by the National Natural Science Foundation of China (NSFC) under
the Grant No. 11535012. JYP is supported in part from the DFG (CRC
110 ``Symmetries and the Emergence of Structure in QCD''). XNW is
supported in part by the National Natural Science Foundation of China
(NSFC) under the Grant No. 11221504 and by the Chinese Ministry of
Science and Technology under Grant No. 2014DFG02050, and by the Director,
Office of Energy Research, Office of High Energy and Nuclear Physics,
Division of Nuclear Physics, of the U.S. Department of Energy under
Contract No. DE- AC02-05CH11231. QW thanks the hospitality of Frankfurt
Institute for Advanced Studies (FIAS) and Institute for Theoretical
Physics at Goethe University Frankfurt where the work was completed. 

\appendix

\section{Derivation of Eq. (\ref{eq:d-j5})}

\label{sec:derivation-d-j5}Let us treat the vorticity term in Eq.
(\ref{eq:d-j5}) as, 
\begin{eqnarray}
I_{\omega} & = & -\frac{1}{2}\hbar\beta\tilde{\Omega}^{\mu\sigma}\partial_{\mu}(\beta\mu)\int d^{4}pp_{\sigma}V_{\beta p_{0},\beta\mu}^{\prime\prime}\delta(p^{2}-m^{2})\nonumber \\
 &  & -\frac{1}{2}\hbar\beta\tilde{\Omega}^{\mu\sigma}\partial_{\mu}(\beta u_{\rho})\int d^{4}pp_{\sigma}p^{\rho}V_{\beta p_{0},\beta p_{0}}^{\prime\prime}\delta(p^{2}-m^{2})\nonumber \\
 & = & \frac{1}{2}\hbar\beta^{2}Q(E\cdot\omega)\int d^{4}pp_{0}V_{\beta p_{0},\beta\mu}^{\prime\prime}\delta(p^{2}-m^{2}),\label{eq:i2}
\end{eqnarray}
where the second term in the first equality is vanishing. This can
be seen by 
\begin{eqnarray}
 &  & \tilde{\Omega}^{\mu\sigma}\partial_{\mu}(\beta u_{\rho})\int d^{4}pp_{\sigma}p^{\rho}V_{\beta p_{0},\beta p_{0}}^{\prime\prime}\delta(p^{2}-m^{2})\nonumber \\
 & = & \tilde{\Omega}^{\mu\sigma}\partial_{\mu}(\beta u_{\rho})u^{\rho}u_{\sigma}\int d^{4}pp_{0}^{2}V_{\beta p_{0},\beta p_{0}}^{\prime\prime}\delta(p^{2}-m^{2})\nonumber \\
 &  & +\frac{1}{3}\tilde{\Omega}^{\mu\sigma}\partial_{\mu}(\beta u_{\rho})\Delta_{\sigma}^{\rho}\int d^{4}p\bar{p}^{2}V_{\beta p_{0},\beta p_{0}}^{\prime\prime}\delta(p^{2}-m^{2})\nonumber \\
 & = & 0,
\end{eqnarray}
where we have assumed that $\beta$ is constant and used $\Delta_{\sigma}^{\rho}=g_{\sigma}^{\rho}-u_{\sigma}u^{\rho}$,
$u^{\rho}\partial_{\mu}u_{\rho}=0$ and 
\begin{equation}
\tilde{\Omega}^{\mu\sigma}\partial_{\mu}(\beta u_{\rho})\Delta_{\sigma}^{\rho}=\beta\tilde{\Omega}^{\mu\sigma}\partial_{\mu}u_{\sigma}=2\beta\partial_{\alpha}\omega^{\alpha}=0.
\end{equation}

Then we look at the second term in Eq. (\ref{eq:d-j5}) which is related
to electromagnetic field, 
\begin{eqnarray}
I_{F} & = & \hbar\beta Q\tilde{F}^{\mu\lambda}\int d^{4}p[-p_{\lambda}(\partial_{\mu}\mu)+(\partial_{\mu}u_{\rho})p_{\lambda}p^{\rho}]V_{\beta p_{0}}^{\prime}\delta^{\prime}(p^{2}-m^{2})\nonumber \\
 & = & \hbar\beta Q^{2}(E\cdot B)\int d^{4}pp_{0}V_{\beta p_{0}}^{\prime}\delta^{\prime}(p^{2}-m^{2})\nonumber \\
 &  & +\frac{1}{3}\hbar\beta Q\tilde{F}^{\mu\lambda}(\partial_{\mu}u_{\lambda})\int d^{4}p\bar{p}^{2}V_{\beta p_{0}}^{\prime}\delta^{\prime}(p^{2}-m^{2})\nonumber \\
 & = & \hbar\beta Q^{2}(E\cdot B)\int d^{4}pp_{0}V_{\beta p_{0}}^{\prime}\delta^{\prime}(p^{2}-m^{2})\nonumber \\
 &  & +\frac{2}{3}\hbar\beta Q(E\cdot\omega)\int d^{4}p\bar{p}^{2}V_{\beta p_{0}}^{\prime}\delta^{\prime}(p^{2}-m^{2})\nonumber \\
 & = & \frac{1}{2}\hbar Q^{2}\beta^{2}(E\cdot B)\int d^{4}pV_{\beta\mu,\beta p_{0}}^{\prime\prime}\delta(p^{2}-m^{2})\nonumber \\
 &  & +\hbar Q\beta(\omega\cdot E)\int d^{4}p\frac{1}{3p_{0}^{2}}\bar{p}^{2}V_{\beta p_{0}}^{\prime}\delta(p^{2}-m^{2})\nonumber \\
 &  & -\hbar Q\beta^{2}(\omega\cdot E)\int d^{4}p\frac{1}{3p_{0}}\bar{p}^{2}V_{\beta p_{0},\beta p_{0}}^{\prime\prime}\delta(p^{2}-m^{2}),\label{eq:i3}
\end{eqnarray}
 where we have used $\delta^{\prime}(x)=-\delta(x)/x$, $\delta^{\prime}(p^{2}-m^{2})\equiv d\delta(p^{2}-m^{2})/dp_{0}^{2}$,
$\partial_{\mu}\mu=-QE_{\mu}$, and 
\begin{eqnarray}
\partial_{\mu}V & = & \beta[(\partial_{\mu}\mu)V_{\beta\mu}^{\prime}+(\partial_{\mu}u_{\rho})p^{\rho}V_{\beta p_{0}}^{\prime}]\nonumber \\
\frac{d}{dp_{0}}(\partial_{\mu}V) & = & \frac{d}{dp_{0}}\left[\partial_{\mu}(\beta\mu)V_{\beta\mu}^{\prime}+\partial_{\mu}(\beta u_{\rho})p^{\rho}V_{\beta p_{0}}^{\prime}\right]\nonumber \\
 & = & \beta\partial_{\mu}(\beta\mu)V_{\beta\mu,\beta p_{0}}^{\prime\prime}+\partial_{\mu}(\beta u_{\rho})u^{\rho}V_{\beta p_{0}}^{\prime}+\beta\partial_{\mu}(\beta u_{\rho})p^{\rho}V_{\beta p_{0},\beta p_{0}}^{\prime\prime}\nonumber \\
 & = & \beta^{2}\left[(\partial_{\mu}\mu)V_{\beta\mu,\beta p_{0}}^{\prime\prime}+(\partial_{\mu}u_{\rho})p^{\rho}V_{\beta p_{0},\beta p_{0}}^{\prime\prime}\right]\nonumber \\
(\partial_{\mu}u_{\lambda})\tilde{F}^{\mu\lambda} & = & \frac{1}{2}\epsilon^{\mu\lambda\alpha\beta}(\partial_{\mu}u_{\lambda})F_{\alpha\beta}\nonumber \\
 & = & \frac{1}{2}\epsilon^{\mu\lambda\alpha\beta}(\partial_{\mu}u_{\lambda})(E_{\alpha}u_{\beta}-E_{\beta}u_{\alpha}+\epsilon_{\alpha\beta\rho\sigma}u^{\rho}B^{\sigma})\nonumber \\
 & = & 2\omega\cdot E+\frac{1}{2}\epsilon^{\mu\lambda\alpha\beta}\epsilon_{\alpha\beta\rho\sigma}(\partial_{\mu}u_{\lambda})u^{\rho}B^{\sigma}\nonumber \\
 & = & 2\omega\cdot E-(\partial_{\rho}u_{\sigma}-\partial_{\sigma}u_{\rho})u^{\rho}B^{\sigma}\nonumber \\
 & = & 2\omega\cdot E.
\end{eqnarray}
We can add $I_{\omega}$ and $I_{F}$ from Eqs. (\ref{eq:i2},\ref{eq:i3})
to obtain the right-hand side of Eq. (\ref{eq:i23-2}), which we denote
as $I=I_{\omega}+I_{F}=I_{E\cdot B}+I_{E\cdot\omega}$ where $I_{E\cdot B}$
and $I_{E\cdot\omega}$ denote the $E\cdot B$ and $E\cdot\omega$
terms respectively. We now work on $I_{E\cdot B}$ and $I_{E\cdot\omega}$.
We now evaluate $I_{E\cdot B}$ as, 
\begin{eqnarray}
I_{E\cdot B} & = & -\frac{1}{2}\hbar Q^{2}\beta^{2}(E\cdot B)\int d^{4}pV_{\beta p_{0},\beta p_{0}}^{\prime\prime}\delta(p^{2}-m^{2})\nonumber \\
 & = & \hbar Q^{2}\beta^{2}(E\cdot B)\int\frac{d^{3}p}{(2\pi)^{3}}\frac{1}{E_{p}}\nonumber \\
 &  & \frac{d}{d(\beta E_{p})}\left\{ f_{\mathrm{FD}}(E_{p}-\mu)[1-f_{\mathrm{FD}}(E_{p}-\mu)]+f_{\mathrm{FD}}(E_{p}+\mu)[1-f_{\mathrm{FD}}(E_{p}+\mu)]\right\} \nonumber \\
 & = & -\hbar Q^{2}\beta(E\cdot B)\int\frac{d^{3}p}{(2\pi)^{3}}\frac{1}{E_{p}^{2}-m\text{\texttwosuperior}}\nonumber \\
 &  & \left\{ f_{\mathrm{FD}}(E_{p}-\mu)[1-f_{\mathrm{FD}}(E_{p}-\mu)]+f_{\mathrm{FD}}(E_{p}+\mu)[1-f_{\mathrm{FD}}(E_{p}+\mu)]\right\} \nonumber \\
 & = & -\frac{1}{2\pi^{2}}\hbar Q^{2}(E\cdot B)C_{1}(\beta m,\beta\mu),\label{eq:anomaly}
\end{eqnarray}
where we have used $d^{3}p=dE_{p}E_{p}\sqrt{E_{p}^{2}-m\text{\texttwosuperior}}$
and $C_{1}(\beta m,\beta\mu)$ is given in Eq. (\ref{eq:c1c2}). Then
the result of $I_{E\cdot\omega}$ is 
\begin{eqnarray}
I_{E\cdot\omega} & = & \frac{1}{2}\hbar\beta Q(E\cdot\omega)\int d^{4}p\left[\frac{2\bar{p}^{2}}{3p_{0}^{2}}V_{\beta p_{0}}^{\prime}-\beta\left(p_{0}+\frac{2\bar{p}^{2}}{3p_{0}}\right)V_{\beta p_{0},\beta p_{0}}^{\prime\prime}\right]\delta(p^{2}-m^{2})\nonumber \\
 & = & \frac{2}{3}\hbar\beta Q(E\cdot\omega)\int\frac{d^{3}p}{(2\pi)^{3}E_{p}}\left(1-\frac{m^{2}}{E_{p}^{2}}\right)\nonumber \\
 &  & \left\{ f_{\mathrm{FD}}(E_{p}-\mu)[1-f_{\mathrm{FD}}(E_{p}-\mu)]-f_{\mathrm{FD}}(E_{p}+\mu)[1-f_{\mathrm{FD}}(E_{p}+\mu)]\right\} \nonumber \\
 &  & +\frac{1}{3}\hbar\beta^{2}Q(E\cdot\omega)\int\frac{d^{3}p}{(2\pi)^{3}}\left(1+\frac{2m^{2}}{E_{p}^{2}}\right)\nonumber \\
 &  & \left\{ [f_{\mathrm{FD}}(E_{p}-\mu)][1-f_{\mathrm{FD}}(E_{p}-\mu)][2f_{\mathrm{FD}}(E_{p}-\mu)-1]\right.\nonumber \\
 &  & \left.-[f_{\mathrm{FD}}(E_{p}+\mu)][1-f_{\mathrm{FD}}(E_{p}+\mu)][2f_{\mathrm{FD}}(E_{p}+\mu)-1]\right\} .\label{eq:i23-1}
\end{eqnarray}
The last term in Eq. (\ref{eq:i23-1}) can be further simplified by
using the formula and the integral by part for $E_{p}$, 
\begin{eqnarray}
\frac{d}{d(\beta x)}\left\{ f_{\mathrm{FD}}(x)[1-f_{\mathrm{FD}}(x)]\right\}  & = & f_{\mathrm{FD}}(x)[1-f_{\mathrm{FD}}(x)][2f_{\mathrm{FD}}(x)-1],\nonumber \\
\frac{d}{dE_{p}}\left[E_{p}\sqrt{E_{p}^{2}-m\text{\texttwosuperior}}\left(1+\frac{2m^{2}}{E_{p}^{2}}\right)\right] & = & E_{p}\sqrt{E_{p}^{2}-m\text{\texttwosuperior}}\left[\left(\frac{1}{E_{p}}+\frac{E_{p}}{E_{p}^{2}-m\text{\texttwosuperior}}\right)\left(1+\frac{2m^{2}}{E_{p}^{2}}\right)-4\frac{m^{2}}{E_{p}^{3}}\right].
\end{eqnarray}
Then Eq. (\ref{eq:i23-1}) can be further simplified as 
\begin{eqnarray}
I_{E\cdot\omega} & = & -\hbar\beta Q(E\cdot\omega)\int\frac{d^{3}p}{(2\pi)^{3}E_{p}}\cdot\frac{m^{2}}{E_{p}^{2}-m^{2}}\nonumber \\
 &  & \left\{ f_{\mathrm{FD}}(E_{p}-\mu)[1-f_{\mathrm{FD}}(E_{p}-\mu)]-f_{\mathrm{FD}}(E_{p}+\mu)[1-f_{\mathrm{FD}}(E_{p}+\mu)]\right\} \nonumber \\
 & = & -\frac{m^{2}}{2\pi^{2}}\hbar Q\beta(E\cdot\omega)C_{2}(\beta m,\beta\mu),\label{eq:final-i23-1}
\end{eqnarray}
where $C_{2}(\beta m,\beta\mu)$ is given by Eq. (\ref{eq:c1c2}).
The right-hand side of Eq. (\ref{eq:d-j5}) is just $I_{E\cdot B}+I_{E\cdot\omega}$,
where $I_{E\cdot B}$ and $I_{E\cdot\omega}$ are in Eqs. (\ref{eq:anomaly},\ref{eq:final-i23-1})
respectively, which finally gives Eq. (\ref{eq:i23-final}).

\section{Small mass expansion of $C_{1}(\beta m,\beta\mu)$ and $C_{2}(\beta m,\beta\mu)$}

\label{sec:c1c2}In this appendix, we expand $C_{1}(\beta m,\beta\mu)$
and $C_{2}(\beta m,\beta\mu)$ in small $\beta m$. For simplicity
of notations, we use new variables $\bar{\mu}\equiv\beta\mu$, $\bar{m}\equiv\beta m$
and define two dimensionless functions 
\begin{eqnarray}
h(\bar{m},\bar{\mu}) & = & \int_{0}^{\infty}dx\frac{1}{e^{\sqrt{x^{2}+\bar{m}^{2}}+\bar{\mu}}+1},\nonumber \\
g(\bar{m},\bar{\mu}) & = & \int_{0}^{\infty}dx\frac{1}{\sqrt{x^{2}+\bar{m}^{2}}}\frac{1}{e^{\sqrt{x^{2}+\bar{m}^{2}}+\bar{\mu}}+1}.
\end{eqnarray}
We can express $C_{1}(\bar{m},\bar{\mu})$ and $C_{2}(\bar{m},\bar{\mu})$
in terms of $h(\bar{m},\bar{\mu})$ and $g(\bar{m},\bar{\mu})$, 

\begin{align}
C_{1}(\bar{m},\bar{\mu}) & =\frac{d}{d\bar{\mu}}[h(\bar{m},-\bar{\mu})-h(\bar{m},\bar{\mu})],\nonumber \\
C_{2}(\bar{m},\bar{\mu}) & =\frac{d}{d\bar{\mu}}[g(\bar{m},-\bar{\mu})+g(\bar{m},\bar{\mu})].
\end{align}
The Fermi-Dirac distribution can be expanded as 
\begin{equation}
\frac{1}{e^{\sqrt{x^{2}+\bar{m}^{2}}+\bar{\mu}}+1}=\sum_{n=1}^{\infty}(-1)^{n+1}e^{-n(\sqrt{x^{2}+\bar{m}^{2}}+\bar{\mu})},
\end{equation}
and then we have 
\begin{eqnarray}
h(\bar{m},\bar{\mu}) & = & \sum_{n=1}^{\infty}(-1)^{n+1}e^{-n\bar{\mu}}\int_{0}^{\infty}dxe^{-n\sqrt{x^{2}+\bar{m}^{2}}}\nonumber \\
 & = & \bar{m}\sum_{n=1}^{\infty}(-1)^{n+1}e^{-n\bar{\mu}}K_{1}(n\bar{m}),\nonumber \\
g(\bar{m},\bar{\mu}) & = & \sum_{n=1}^{\infty}(-1)^{n+1}e^{-n\bar{\mu}}\int_{0}^{\infty}dx\frac{1}{\sqrt{x^{2}+\bar{m}^{2}}}e^{-n\sqrt{x^{2}+\bar{m}^{2}}}\nonumber \\
 & = & \sum_{n=1}^{\infty}(-1)^{n+1}e^{-n\bar{\mu}}K_{0}(n\bar{m}),
\end{eqnarray}
where $K_{0}(x)$ and $K_{1}(x)$ are modified Bessel functions of
the second kind whose expansion form are 
\begin{eqnarray}
K_{0}(x) & = & -(\ln\frac{x}{2}+\gamma)\sum_{k=0}^{\infty}\frac{1}{(k!)^{2}}\left(\frac{x}{2}\right)^{2k}+\sum_{k=1}^{\infty}\frac{1}{(k!)^{2}}\left(1+\frac{1}{2}+\cdots+\frac{1}{k}\right)\left(\frac{x}{2}\right)^{2k},\nonumber \\
K_{1}(x) & = & \frac{1}{x}+\left(\ln\frac{x}{2}+\gamma\right)\sum_{k=0}^{\infty}\frac{1}{k!(k+1)!}\left(\frac{x}{2}\right)^{2k+1}\nonumber \\
 &  & -\frac{1}{2}\sum_{k=0}^{\infty}\frac{1}{k!(k+1)!}\left(1+\frac{1}{2}+\cdots+\frac{1}{k+1}\right)\left(\frac{x}{2}\right)^{2k+1}\nonumber \\
 &  & -\frac{1}{2}\sum_{k=0}^{\infty}\frac{1}{(k+1)!(k+2)!}\left(1+\frac{1}{2}+\cdots+\frac{1}{k+1}\right)\left(\frac{x}{2}\right)^{2k+3},
\end{eqnarray}
with the Euler constant $\gamma=0.577\cdots$. 

Then one can verify that $h(\bar{m},\bar{\mu})$ can be cast into
the following form

\begin{eqnarray}
h(\bar{m},\bar{\mu}) & = & \phi(\bar{\mu})-\bar{m}\sum_{k=0}^{\infty}\frac{1}{k!(k+1)!}\left(\frac{\bar{m}}{2}\frac{d}{d\bar{\mu}}\right)^{2k+1}\left[\chi(\bar{\mu})+\frac{1}{e^{\bar{\mu}}+1}\ln\frac{\bar{m}e^{\gamma}}{2}\right]\nonumber \\
 &  & +\frac{\bar{m}}{2}\sum_{k=0}^{\infty}\frac{1}{k!(k+1)!}\left(1+\frac{1}{2}+\cdots+\frac{1}{k+1}\right)\left(\frac{\bar{m}}{2}\frac{d}{d\bar{\mu}}\right)^{2k+1}\frac{1}{e^{\bar{\mu}}+1}\nonumber \\
 &  & +\frac{\bar{m}}{2}\sum_{k=0}^{\infty}\frac{1}{(k+1)!(k+2)!}\left(1+\frac{1}{2}+\cdots+\frac{1}{k+1}\right)\left(\frac{\bar{m}}{2}\frac{d}{d\bar{\mu}}\right)^{2k+3}\frac{1}{e^{\bar{\mu}}+1},
\end{eqnarray}
where we have defined two series 
\begin{eqnarray}
\chi(\bar{\mu}) & = & \sum_{n=1}^{\infty}(-1)^{n+1}e^{-n\bar{\mu}}\ln n,\nonumber \\
\phi(\bar{\mu}) & = & \sum_{n=1}^{\infty}(-1)^{n+1}e^{-n\bar{\mu}}\frac{1}{n}.
\end{eqnarray}
Obviously we have following relation for $\phi(\bar{\mu})$, 
\begin{eqnarray}
\frac{d}{d\bar{\mu}}\phi(\bar{\mu}) & = & -\sum_{n=1}^{\infty}(-1)^{n+1}e^{-n\bar{\mu}}=-\frac{1}{e^{\bar{\mu}}+1},\nonumber \\
\frac{d}{d\bar{\mu}}\phi(-\bar{\mu}) & = & -\frac{d}{d(-\bar{\mu})}\phi(-\bar{\mu})=\frac{e^{\bar{\mu}}}{e^{\bar{\mu}}+1}.
\end{eqnarray}
Then we obtain 
\begin{eqnarray}
h(\bar{m},-\bar{\mu})-h(\bar{m},\bar{\mu}) & = & \phi(-\bar{\mu})-\phi(\bar{\mu})+\bar{m}I_{1}\left(\bar{m}\frac{d}{d\bar{\mu}}\right)\left[\chi(\bar{\mu})+\chi(-\bar{\mu})\right],\nonumber \\
\frac{d}{d\bar{\mu}}\left[h(\bar{m},-\bar{\mu})-h(\bar{m},\bar{\mu})\right] & = & 1+\bar{m}I_{1}\left(\bar{m}\frac{d}{d\bar{\mu}}\right)\frac{d}{d\bar{\mu}}\left[\chi(\bar{\mu})+\chi(-\bar{\mu})\right],
\end{eqnarray}
with the modified Bessel function $I_{1}(x)$ of the first kind is
given by 
\begin{equation}
I_{1}(x)=\sum_{k=0}^{\infty}\frac{1}{k!(k+1)!}\bigg(\frac{x}{2}\bigg)^{2k+1}.
\end{equation}

We can also treat $g(\bar{m},\bar{\mu})$ in the same way 
\begin{eqnarray}
g(\bar{m},\bar{\mu}) & = & -\sum_{k=0}^{\infty}\frac{1}{(k!)^{2}}\left(\frac{\bar{m}}{2}\frac{d}{d\bar{\mu}}\right)^{2k}\left[\chi(\bar{\mu})+\frac{1}{e^{\bar{\mu}}+1}\ln\frac{\bar{m}e^{\gamma}}{2}\right]\nonumber \\
 &  & +\sum_{k=1}^{\infty}\frac{1}{(k!)^{2}}\left(1+\frac{1}{2}+\cdots+\frac{1}{k}\right)\left(\frac{\bar{m}}{2}\frac{d}{d\bar{\mu}}\right)^{2k}\frac{1}{e^{\bar{\mu}}+1},
\end{eqnarray}
and we obtain 
\begin{eqnarray}
g(\bar{m},-\bar{\mu})+g(\bar{m},\bar{\mu}) & = & -\ln\frac{\bar{m}e^{\gamma}}{2}-I_{0}\left(\bar{m}\frac{d}{d\bar{\mu}}\right)\left[\chi(\bar{\mu})+\chi(-\bar{\mu})\right],\nonumber \\
\frac{d}{d\bar{\mu}}\left[g(\bar{m},-\bar{\mu})+g(\bar{m},\bar{\mu})\right] & = & -I_{0}\left(\bar{m}\frac{d}{d\bar{\mu}}\right)\frac{d}{d\bar{\mu}}\left[\chi(\bar{\mu})+\chi(-\bar{\mu})\right],
\end{eqnarray}
with the modified Bessel function $I_{0}(x)$ of the first kind is
given by 
\begin{equation}
I_{0}(x)=\sum_{k=0}^{\infty}\frac{1}{(k!)^{2}}\bigg(\frac{x}{2}\bigg)^{2k}.
\end{equation}

Finally we obtain the expressions for $C_{1}(\bar{m},\bar{\mu})$
and $C_{2}(\bar{m},\bar{\mu})$, 
\begin{align}
C_{1}(\bar{m},\bar{\mu}) & =1+\bar{m}I_{1}\left(\bar{m}\frac{d}{d\bar{\mu}}\right)\frac{d}{d\bar{\mu}}\left[\chi(\bar{\mu})+\chi(-\bar{\mu})\right],\nonumber \\
C_{2}(\bar{m},\bar{\mu}) & =-I_{0}\left(\bar{m}\frac{d}{d\bar{\mu}}\right)\frac{d}{d\bar{\mu}}\left[\chi(\bar{\mu})+\chi(-\bar{\mu})\right].
\end{align}
The series $\chi(\bar{\mu})$ can also be written in the form, 
\begin{eqnarray}
\chi(\bar{\mu}) & = & \sum_{n=1}^{\infty}(-1)^{n+1}e^{-n\bar{\mu}}\ln n=\left[\frac{\partial}{\partial y}\sum_{n=1}^{\infty}\frac{(-e^{-\bar{\mu}})^{n}}{n^{y}}\right]_{y=0}\nonumber \\
 & = & \left[\frac{\partial}{\partial y}\text{Li}(y,-e^{-\bar{\mu}})\right]_{y=0},
\end{eqnarray}
where $\text{Li}(y,z)$ is the polylogarithm function,
\begin{equation}
\text{Li}(y,z)=\sum_{n=1}^{\infty}\frac{z^{n}}{n^{y}},\ \ |z|<1.
\end{equation}
The analytic continuation of the polylogarithm beyond the circle of
convergence $|z|<1$ can be furnished by following integral representation,
\begin{equation}
\text{Li}(y,z)=\frac{1}{\Gamma(y)}\int_{0}^{\infty}\frac{t^{y-1}}{z^{-1}e^{t}-1}dt,\ \ \text{Re}(y)>0,\: z\in\mathbb{C}/[1,\infty),
\end{equation}
with the Gamma function $\Gamma(y)$ defined by 
\begin{equation}
\Gamma(y)=\int_{0}^{\infty}e^{-t}t^{y-1}dt,\ \ \text{Re}(y)>0.
\end{equation}
The asymptotic behaviors of $I_{0}(x)$ and $I_{1}(x)$ at $x\rightarrow0$
are $I_{0}(x)\approx1$ and $I_{1}(x)\approx x/2$. So at $\bar{m}\rightarrow0$,
we obtain the asymptotic values of $C_{1}(\bar{m},\bar{\mu})$ and
$C_{2}(\bar{m},\bar{\mu})$, 
\begin{align}
C_{1}(\bar{m},\bar{\mu}) & \approx1+\frac{\bar{m}^{2}}{2}\frac{d^{2}}{d\bar{\mu}^{2}}\frac{d}{dy}\left[\text{Li}(y,-e^{-\bar{\mu}})+\text{Li}(y,-e^{\bar{\mu}})\right]_{y=0},\nonumber \\
C_{2}(\bar{m},\bar{\mu}) & \approx-\frac{d}{d\bar{\mu}}\frac{d}{dy}\left[\text{Li}(y,-e^{-\bar{\mu}})+\text{Li}(y,-e^{\bar{\mu}})\right]_{y=0}.\label{eq:asymptotic-c1c2}
\end{align}

\section{Derivation of Eq. (\ref{eq:p-derivative})}

\label{sec:p-derivative}In this appendix, we give a detailed derivation
of Eq. (\ref{eq:p-derivative}). From the definition of the Wigner
function (\ref{eq:wigner-def-1}) and that of the axial vector component,
we obtain

\begin{eqnarray}
QF_{\mu\nu}\int d^{4}p\partial_{p}^{\nu}\mathscr{A}^{\mu} & = & QF_{\mu\nu}\int d^{4}p\partial_{p}^{\nu}\int\frac{d^{4}y}{(2\pi)^{4}}e^{-ip\cdot y}\nonumber \\
 &  & \times\left\langle \bar{\psi}(x+\frac{1}{2}y)\gamma^{\mu}\gamma^{5}\mathrm{P}U(G,x+\frac{1}{2}y,x-\frac{1}{2}y)\psi(x-\frac{1}{2}y)\right\rangle \nonumber \\
 & = & QF_{\mu\nu}\int d^{4}y(-iy^{\nu})\delta^{(4)}(y)\nonumber \\
 &  & \times\left\langle \bar{\psi}(x+\frac{1}{2}y)\gamma^{\mu}\gamma^{5}\mathrm{P}U(G,x+\frac{1}{2}y,x-\frac{1}{2}y)\psi(x-\frac{1}{2}y)\right\rangle \nonumber \\
 & = & -iQF_{\mu\nu}\bigg(\lim_{y\rightarrow0}y^{\nu}\left\langle \bar{\psi}(x+\frac{y}{2})\gamma^{\mu}\gamma^{5}\psi(x-\frac{y}{2})\right\rangle \bigg)\nonumber \\
 & = & -\frac{Q^{2}}{8\pi^{2}}F_{\mu\nu}\tilde{F}^{\mu\nu}\nonumber \\
 & = & -\frac{Q^{2}}{2\pi^{2}}B\cdot E
\end{eqnarray}
where we have used $4B\cdot E=F_{\mu\nu}\tilde{F}^{\mu\nu}$, $\lim_{y\rightarrow0}\mathrm{P}U(G,x+\frac{1}{2}y,x-\frac{1}{2}y)=1$
and \cite{Peskin} 
\begin{eqnarray}
\lim_{y\rightarrow0}\bar{\psi}(x+\frac{y}{2})\gamma^{\mu}\gamma^{5}\psi(x-\frac{y}{2}) & = & -\frac{i}{4\pi^{2}}Q\epsilon^{\alpha\beta\mu\rho}F_{\alpha\beta}\lim_{y\rightarrow0}\frac{y_{\rho}}{y^{2}}\nonumber \\
\lim_{y\rightarrow0}\frac{y^{\nu}y^{\rho}}{y^{2}} & = & \frac{1}{4}g^{\nu\rho}
\end{eqnarray}

\bibliographystyle{apsrev}
\addcontentsline{toc}{section}{\refname}\bibliography{ref-1}

\end{document}